\documentstyle[aps,prl]{revtex}

\begin{document}

\title{Conditional Statistics of Temperature Fluctuations in Turbulent Convection}

\author{Emily S.C. Ching and K.L. Chau}
\address{Department of Physics, The Chinese University of Hong Kong,
Shatin, Hong Kong}

\date{\today}

\maketitle

\begin{abstract}

We find that the conditional statistics of temperature difference
at fixed values of the locally averaged temperature dissipation rate
in turbulent convection become Gaussian in the regime where the 
mixing dynamics is expected to be driven by buoyancy. Hence, 
intermittency of the temperature fluctuations in this buoyancy-driven 
regime can be solely attributed to the variation of the locally 
averaged temperature dissipation rate. We further obtain the functional 
behavior of these conditional temperature structure functions. This
functional form demonstrates explicitly the failure of dimensional agruments
and enhances the understanding of the temperature structure functions.

\end{abstract}

\pacs{PACS numbers: 47.27.-i, 05.40.-a }

In turbulent fluid flows, physical quantities such as velocity,
temperature and pressure exhibit seemingly irregular fluctuations 
both in time and in space. A key issue in turbulence research is to make sense
of these fluctuations. The central result of the seminal work of Kolmogorov
in 1941 (K41)\cite{K41} is that the fluctuating velocity field in
high Reynolds number Navier-Stokes turbulence is self-similar at scales
within the inertial range, the range of length scales that are smaller than
those of energy input and larger than those affected directly by molecular
dissipation. K41 predicted that the velocity structure functions 
$\langle [u(x+r)-u(x)]^p \rangle$ scale as $r^{\xi_p}$ with scaling
exponents $\xi_p$ equal to $p/3$ when $r$ is within the inertial range.
Experimental and numerical results, however, indicate that
$\xi_p$ is a nonlinear function of $p$ and that turbulent 
velocity fluctuations are scale-dependent in that the shape
of the probability density function (PDF) of the velocity difference 
$u(x+r)-u(x)$ changes with the scale $r$ even when $r$ is within
the inertial range. This deviation from the K41 results is associated with
intermittency or the uneven distribution of turbulent activity of
the velocity field in time and in space.

Extensive efforts have been devoted to the understanding of 
the problem of intermittency or anomalous scaling.
In his Refined Similarity Hypothesis (RSH)\cite{K62,O62},
Kolmogorov attributed this intermittent nature of the velocity fluctuations 
to the spatial variations of the energy dissipation rate.
Various models have been put forth for the statistics of the locally averaged 
energy dissipation rate. The most recent model of She and Leveque\cite{SL94}
proposed a hierarchical structure for the moments, which 
leads to predictions that are 
in good agreement with experiments. This moment hierarchy was later shown
to be naturally satisfied by log-Poisson statistics\cite{D94,SW95}.

High Rayleigh number convection has been a well-studied model system for 
investigating turbulence. Fluid motion is driven by an applied temperature 
difference across the top and the bottom plates of a closed experimental cell 
filled with fluid. The temperature field in convection is thus an so-called 
active scalar. The flow state is characterized by the geometry
of the cell and two dimensionless parameters: the Rayleigh number 
Ra $ = \alpha g \Delta L^3 / (\nu \kappa)$
and the Prandtl number  Pr $ = \nu/\kappa$,
where $L$ is the height of the cell, $\Delta$ is the applied
temperature difference, $g$ the acceleration due to gravity,
and $\alpha$, $\nu$, and $\kappa$ are respectively the volume
expansion coefficient, the kinematic viscosity and the thermal
diffusivity of the fluid. When Ra is large enough,
the convection becomes turbulent. 

In turbulent convection, the temperature fluctuations are also 
intermittent\cite{Ching91}. As for velocity fluctuations in high Reynolds 
number Navier-Stokes turbulence, it is of interest to understand the 
intermittency of temperature fluctuations in high Rayleigh number
convection. Turbulent convection poses additional interesting questions 
of its own. There is the issue of whether and how the characteristics 
of turbulence are affected by the presence of buoyancy. One expects 
the mixing dynamics to be driven by buoyancy at scales larger than the 
Bolgiano scale, 
$l_B \equiv {\bar \epsilon}^{5/4} / [{\bar \chi}^{3/4} 
(\alpha g)^{3/2}]$\cite{Monin},
where $\bar \epsilon$ and $\bar \chi$ are respectively the average energy and 
temperature (variance) dissipation rates. On the other hand, for length 
scales smaller than $l_B$, the mixing dynamics is expected to be
driven by the inertial force of the fluid motion and the temperature
field is effectively passive. Recently, one of us (Ching)\cite{Ching00} 
has analyzed the intermittency of temperature field in turbulent 
convection. The normalized temperature structure functions 
have indeed been found to have different scaling exponents 
in the buoyancy-driven and in the 
inertia-driven regimes.

In our present project, we attempt to understand the
intermittency problem of temperature by separating it
into two parts: the understanding of the conditional statistics
of temperature fluctuations at fixed values of the locally averaged
temperature dissipation rate and the understanding of the
statistics of the local temperature dissipation.
In this paper, we report our study of the first part. 
The second part of our study is reported elsewhere\cite{preprint}. 
This separation allows us to especially address whether
RSH type ideas would be fruitful. We shall see that the 
intermittent nature of the temperature fluctuations
in the buoyancy-driven regime 
can indeed be attributed to the variations of the 
locally averaged temperature dissipation rate. Moreover, a change
in the statistical features of the temperature fluctuations
is again observed when the Bolgiano
scale $l_B$ is crossed. This change manifests itself 
as a change
in the behavior of the conditional PDFs of 
the temperature difference at fixed value of the locally 
averaged temperature dissipation rate.
 
We use temperature data obtained 
in the well-documented Chicago experiment of low-temperature 
helium gas\cite{HCL87,SWL89} for our analyses. The experimental 
cell heated from below is cylindrical with a diameter of 20 cm 
and a height of 40 cm. A mean 
circulating flow is present for Ra $\ge 10^8$. 
The temperature at the center of the cell, $T(t)$, 
was measured as a function of time $t$. We 
evaluate the temperature difference between two times: 
$T_\tau(t) \equiv T(t+\tau)-T(t)$. 
The intermittency of the temperature fluctuations is manifested
as a change in the shape of the PDF of $T_\tau$ as $\tau$ 
varies. In our earlier study of this $\tau$-dependence\cite{Ching91},
the dissipative and the circulation time scales, $\tau_d$ and $\tau_c$, 
were identified. A time scale corresponding to $l_B$ is naturally
defined by $\tau_B = \tau_c \, l_B/L$. It was shown\cite{CCIP93} that 
$l_B$ can be written as
\begin{equation}
l_B = {{\rm Nu}^{1 \over 2} L \over ({\rm Ra} \,  {\rm Pr})^{1 \over 4} }
\label{lB}
\end{equation}
where the Nusselt number (Nu) is the heat flux normalized by that
when there was only conduction. Thus, $\tau_B$ can be easily 
evaluated using the measured values of Nu, Ra, and Pr.

The locally averaged temperature dissipation rate $\chi_r$ is the spatial
average of $\kappa |\nabla T|^2$ over a ball of radius $r$. We 
estimate it by $\chi_\tau$, which is defined as
\begin{equation}
\chi_\tau(t) \equiv {1 \over \tau} \int_t^{t+\tau}
{\kappa \over \langle u_c^2 \rangle}
\left({\partial T \over \partial t'}\right)^2 d t'
\label{chir}
\end{equation}
and can be calculated using the one-point temperature
measurements. Here, $\langle u_c^2 \rangle$ is the mean square 
velocity fluctuations at the center of the cell.

We start by investigating the conditional PDF of $T_\tau$, at 
fixed values of $\chi_\tau$. 
We consider those $T_\tau(t)$ whose
corresponding $\ln \chi_\tau(t)$ assumes a certain value 
within a small range $\delta$,
and calculate the conditional PDFs $P(Y_\tau|\chi_\tau)$ where
\begin{equation}
Y_\tau \equiv 
{T_\tau  \over \sqrt{\langle T_\tau^2 | \chi_\tau \rangle} }
\label{ytau}
\end{equation}
As the conditional mean $\langle T_\tau | \chi_\tau\rangle$
is approximately zero, $P(Y_\tau|\chi_\tau)$ is standardized 
with zero mean and unit standard deviation. For a given $\tau$,
$P(Y_\tau|\chi_\tau)$ is found to be independent of $\chi_\tau$
for a range of $\chi_\tau$ that contains most of the data.
The conditional PDFs 
for different values of $\tau$ are plotted in Fig.~1. 
We measure the value of $\chi_\tau$ in units of $\chi \equiv
\kappa \langle (\partial T/\partial t)^2 \rangle/ \langle u_c^2 \rangle$.
In the limit $\tau \to 0$, $\chi_\tau \sim T_\tau^2$, therefore,
the conditional PDF is bimodal for small $\tau$, as seen in the figure.
As $\tau$ increases, $P(Y_\tau|\chi_\tau)$ changes from bimodal to
a function with one maximum and varies with $\tau$. But for larger $\tau$, 
it becomes a standardized Gaussian distribution and is thus independent 
of $\tau$. Such a change in behavior occurs at $\tau \approx \tau_B$. 

Hence, a change in the statistical features of the temperature fluctuations
is again observed as the Bolgiano scale is crossed, demonstrating that
buoyancy does have an effect on the characteristics of turbulence
in convection. Moreover, the physical nature of the presently observed
change is clear. We have the interesting result that
the temperature fluctuations at fixed values of $\chi_\tau$ 
become self-similar and thus non-intermittent in the regime
where the mixing dynamics is expected to be driven by buoyancy.
In other words, intermittency of the temperature fluctuations in
this buoyancy-driven regime can be solely attributed to the
variations of $\chi_\tau$.

In the remaining of this paper, we shall 
obtain the functional dependence of 
the conditional temperature structure functions
$\langle |T_\tau|^p | \chi_\tau \rangle$
on $p$, $\tau$, and $\chi_\tau$. 

It is illuminating to
first work out what functional form is predicted by
simple phenomenology and dimensional agruments.
One expects $T_r$, the temperature difference across a scale $r$,
depends on $r$, $\chi_r$, and $u_r$, the velocity difference
across the same scale $r$. In the inertia-driven regime, $u_r$ is related
to the locally averaged energy dissipation rate $\epsilon_r$ by
$u_r \sim (r \epsilon_r)^{1/3}$ while in the buoyancy-driven
regime, $u_r$ is generated by buoyancy: 
$u_r^2/r \sim \alpha g T_r$. 
Hence, we have
\begin{equation}
T_r \sim  \cases{r^{1/3} \epsilon_r^{-1/6} \chi_r^{1/2} &
$r < l_B$ \cr r^{1/5} \chi_r^{2/5} (\alpha g)^{-1/5} & $r > l_B$} 
\label{dim}
\end{equation}
Equation (\ref{dim})
implies that
\begin{equation}
\langle |T_\tau|^p | \chi_\tau \rangle \sim
\cases{ \langle u_c^2 \rangle^{p/6}
\tau^{p/3} \chi_\tau^{p/2}
\langle {\epsilon_\tau}^{-p/6} | \chi_\tau \rangle &
$\tau < \tau_B$ \cr \langle u_c^2 \rangle^{p/10} \tau^{p/5}
\chi_\tau^{2p/5} (\alpha g)^{-p/5} & $\tau > \tau_B$ }
\label{Predict}
\end{equation}
if $T_\tau$, $\chi_\tau$, and $\epsilon_\tau$
have the same scaling behavior in $\tau$ as the
corresponding quantities with subscript $r$ in $r$ with
$r = \langle u_c^2 \rangle^{1/2} \tau$.

If the variations of $\chi_\tau$ and $\epsilon_\tau$ are both ignored, 
(\ref{Predict}) implies that the temperature frequency power 
spectrum has a scaling $\omega^{-{7 \over 5}}$ for frequency 
$\omega < \omega_B$ and $\omega^{-{5/3}}$ for $\omega > \omega_B$,
where $\omega_B = 2 \pi / \tau_B$. 
The former scaling behavior was reported for
the temperature frequency power spectra measured in 
water\cite{CCIP93} and helium\cite{WKLS90} 
while the latter one was reported for 
that measured in low Pr mercury\cite{CCS95}.

Now we proceed with the analyses.
From the result that $P(Y_\tau|\chi_\tau)$
is independent of $\chi_\tau$, we get
\begin{equation}
\langle |T_\tau|^p | \chi_\tau \rangle = F_p(\tau)
\sigma^p(\tau,\chi_\tau)
\label{xindep}
\end{equation}
where
\begin{equation}
\sigma(\tau,\chi_\tau) \equiv
\sqrt{\langle T_\tau^2 | \chi_\tau \rangle}
\label{sigma}
\end{equation}
By definition, $F_2(\tau) = 1$.
For $\tau > \tau_B$, $P(Y_\tau|\chi_\tau)$ becomes
a standardized Gaussian, thus
\begin{equation}
F_p(\tau > \tau_B) = \sqrt{ 2^p \over \pi}
\Gamma({p+1 \over 2})
\label{Fp2}
\end{equation}
is independent of $\tau$. 
For $\tau_d < \tau < \tau_B$,
we find that $F_p(\tau)$ can be fitted by a power law~(see Fig.~2), that is
\begin{equation}
F_p(\tau) \approx C_p \tau^{\alpha_p} \qquad \tau_d < \tau < \tau_B
\label{Fp1}
\end{equation}
This $\tau$ dependence of $F_p$ echoes that of
$P(Y_\tau | \chi_\tau)$ for $\tau < \tau_B$. Using (\ref{Predict}), 
such dependence 
can be attributed to the additional variation of the local energy 
dissipation rate $\epsilon_\tau$ even when the local temperature 
dissipation rate $\chi_\tau$ is held fixed.
The scaling exponents $\alpha_p$ are plotted in Fig.~3.
Since $\alpha_0 = \alpha_2 = 0$ by definition, $\alpha_p$
has to be a nonlinear function of $p$, as is found.

Next, we analyze the functional dependence of $\sigma$.
We fix $\tau$ and study its dependence on $\chi_\tau$.
When $\tau$ is not too large, $\sigma(\tau,\chi_\tau)$  
indeed scales with $\chi_\tau$ for a range of $\chi_\tau$ that
contains most of the data.  
The scaling exponent $b(\tau)$, however, varies with $\tau$. 
When $\tau$ is large, the data scatter. 
Thus, we have
\begin{equation}
\sigma(\tau,\chi_\tau) = G(\tau) \chi_\tau^{b(\tau)}
\label{Gtau}
\end{equation}
From the relation $\chi_\tau
\sim T_\tau^2$ in the limit of $\tau \to 0$, one gets
$b(\tau) \to 1/2$ as $\tau \to 0$. Indeed, as shown in Fig.~4,
$b(\tau)$ is about $1/2$ for $\tau \le \tau_d$. It then 
crosses over to an approximately linear function of $\ln \tau$,
and has a value of $2/5$ at $\tau \approx \tau_B$. 
This is, therefore, in contrary to the behavior of
$\sigma(\tau,\chi_\tau) \sim \tau^{1/3} \chi_\tau^{1/2}$
and $\sigma(\tau,\chi_\tau) \sim \tau^{1/5} \chi_\tau^{2/5}$
respectively in the inertia-driven ($\tau_d < \tau < \tau_B$) 
and buoyancy-driven regimes ($\tau_B < \tau < \tau_c$)
that simple phenomenology and dimensional agruments would 
predict [see (\ref{Predict})].
In Fig.~5, we plot 
$\sigma(\tau,\chi_\tau)(\chi_\tau/\chi)^{-b(\tau)}$
for various values of $\chi_\tau$. 
The linear fit of $b(\tau)$ in $\ln \tau$ is used for $\tau > \tau_d$.
The data for different values of $\chi_\tau$ collapse to one single
curve, thus confirming (\ref{Gtau}). We take the average of the
data to get an estimate of $G(\tau) \chi^{b(\tau)}$, which is shown
in the inset. It can be fitted by a power law for $\tau > \tau_B$
with an exponent about 0.27.

The temperature structure functions $\langle |T_\tau|^p \rangle$ 
are related to the conditional ones at fixed values 
of $\chi_\tau$ as follows:
\begin{equation}
\langle |T_\tau|^p \rangle = \int_0^\infty 
\langle |T_\tau|^p | \chi_\tau \rangle P_\tau(\chi_\tau) d \chi_\tau
\end{equation}
where $P_\tau(\chi_\tau)$ is the PDF of $\chi_\tau$.
Using (\ref{xindep}) and (\ref{Gtau}), we thus get 
\begin{equation}
\langle |T_\tau|^p \rangle = F_p(\tau) G^p(\tau) 
\langle \chi_\tau^{p b(\tau)} \rangle
\label{struct}
\end{equation}
Equation (\ref{struct}) implies that
the change in the
scaling exponent of the
normalized structure functions $\langle |T_\tau|^p \rangle/
\langle T_\tau^2 \rangle^{p/2}$ observed, when $\tau_B$ is
crossed\cite{Ching00}, is the combined effect of
the change in the $\tau$ dependence of $F_p(\tau)$ and
$\langle \chi_\tau^{p b(\tau)} \rangle$.
The comparison of (\ref{struct}) with data will be
presented elsewhere. 

In summary, we have studied systematically
the conditional statistics of the temperature fluctuations
at fixed values of local temperature dissipation $\chi_\tau$
in turbulent convection.  We have found that such
conditional statistics become self-similar
in the buoyancy-driven regime, demonstrating that
the intermittency of the temperature field in this regime
can be attributed solely to the variations of $\chi_\tau$.
We have worked out the functional behavior of 
the conditional structure functions 
$\langle |T_\tau|^p | \chi_\tau \rangle$. There is indeed
scaling behavior in $\chi_\tau$ but the scaling exponent 
$b(\tau)$ depends on $\tau$, in contrary to 
what simple phenomenology and dimensional agruments might predict.
We emphasize that this $\tau$ dependence demonstrates explicitly 
the failure of dimensional arguments. Together with the knowledge 
of the statistical properties of $\chi_\tau$,
this functional behavior would enable us to better
understand the temperature structure functions.

ESCC would like to acknowledge discussions with T.~Witten.
This work is supported by a grant from the Research Grants Council 
of the Hong Kong Special Administrative Region, China 
(RGC Ref. No. CUHK 4119/98P).

\newpage

\centerline{FIGURE CAPTIONS}

\vspace{15pt}

\noindent
FIG 1. The conditional PDFs $P(Y_\tau|\chi_\tau)$
versus $Y_\tau$ for Ra = $6.0 \times 10^{11}$ 
and $\chi_\tau/\chi = 0.18$ for various values
of $\tau$. $\tau = 8$ (dotted line), 
$\tau = 16$~(dashed line), $\tau=32$~(dot-dashed line),
$\tau=64$~(circles), $\tau=128$~(squares), and 
$\tau=256$~(triangles). 
It can be seen that $P(Y_\tau|\chi_\tau)$ becomes
a standard Gaussian distribution (solid line) for 
$\tau > \tau_B \approx 70$. All times
are in units of the sampling time = 1/409.6~s.
The conditional PDFs are found to be independent of
$\chi_\tau$.

\vspace{12pt}

\noindent
FIG. 2. The logarithm of the 
normalized conditional temperature structure functions
$F_p(\tau) \equiv \langle |T_\tau|^p | \chi_\tau \rangle/
\langle T_\tau^2 | \chi_\tau \rangle^{p/2}$ versus $\ln \tau$
for Ra = $7.3 \times 10^{10}$ and
$\chi_\tau/\chi = 0.43$ for various values of $p$.
The three time scales $\tau_d$, $\tau_B$ and $\tau_c$ are approximately
$6$, $60$ and $1750$ respectively, and are indicated by the dashed lines. All
times are in units of the sampling time = 1/320~s.
$p=0.5$~(circles), $p=1.5$~(diamonds), $p=1.75$~(triangles), 
$p=2.25$~(crosses), $p=2.5$~(squares), and $p=2.75$~(pluses). 
For $\tau_d < \tau < \tau_B$, $F_p(\tau)$ can be
fitted by a power-law $C_p \tau^{\alpha_p}$~(solid lines) and for $\tau> \tau_B$,
it becomes $\sqrt{2^p/\pi} \Gamma((p+1)/2)$~(dot-dashed lines) and is thus independent
of $\tau$.

\vspace{12pt}

\noindent 
FIG. 3. The scaling exponent $\alpha_p$ versus $p$ for Ra = $4.0 \times
10^9$~(circles), Ra = $7.3 \times 10^{10}$~(squares), and 
Ra = $6.0 \times 10^{11}$~(diamonds). 

\vspace{12pt}

FIG. 4. The scaling exponent $b(\tau)$ versus $\ln \tau$ for 
Ra = $4.0 \times 10^9$. The time scales $\tau_d$ and $\tau_B$ 
are approximately $8$ and $50$ respectively, and are 
indicated by the dashed lines. All times are in units of the
sampling time = 1/160.8~s. It can be seen that $b(\tau)$ is close
to $1/2$ for $\tau \le \tau_d$ and can be fitted by a linear function
in $\ln \tau$ (solid line) for $\tau > \tau_d$. Moreover, 
$b(\tau) \approx 2/5$ at $\tau \approx \tau_B$.

\vspace{12pt}

\noindent 
FIG. 5.  $\ln \sigma(\tau,\chi_\tau) (\chi_\tau/\chi)^{-b(\tau)}$ versus
$\ln \tau$ for Ra = $7.3 \times 10^{10}$ for $\chi_\tau/\chi = 0.13$~(circles),
$\chi_\tau/\chi = 0.35$~(squares), and $\chi_\tau/\chi \approx 0.96$~(triangles).
The three sets of data collapse into a single function of $\tau$ ($= G(\tau)
\chi^{b(\tau)})$ confirming (\ref{Gtau}). The times are in units of the
sampling time = 1/320~s while $\sigma$ is in units of the standard deviation
of the temperature fluctuations. Shown in the inset is the average of the
three sets of data~(solid line), which can be fitted by a power law
(dot-dashed line) for $\tau > \tau_B$~(indicated by dashed line).

\vspace{12pt}

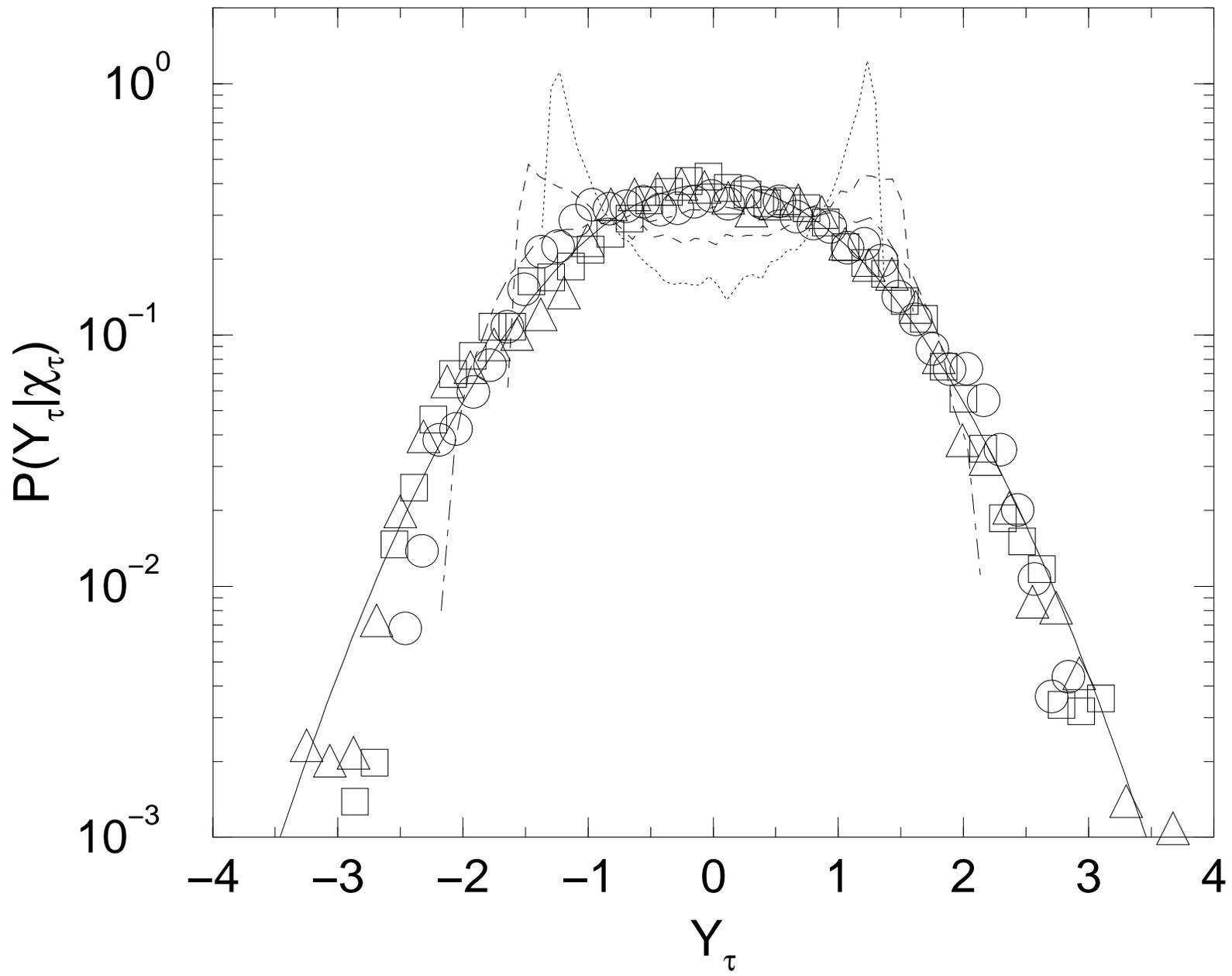
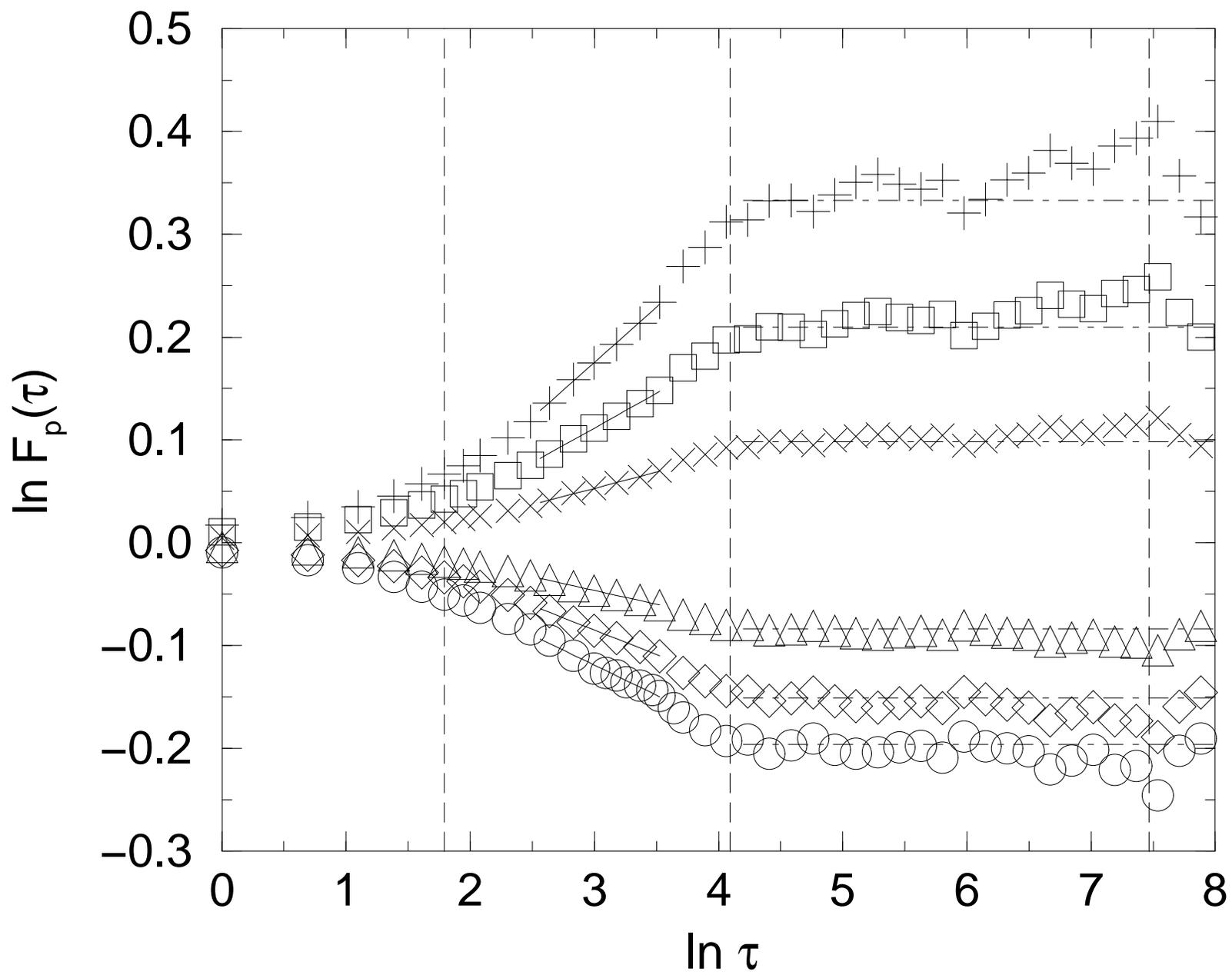
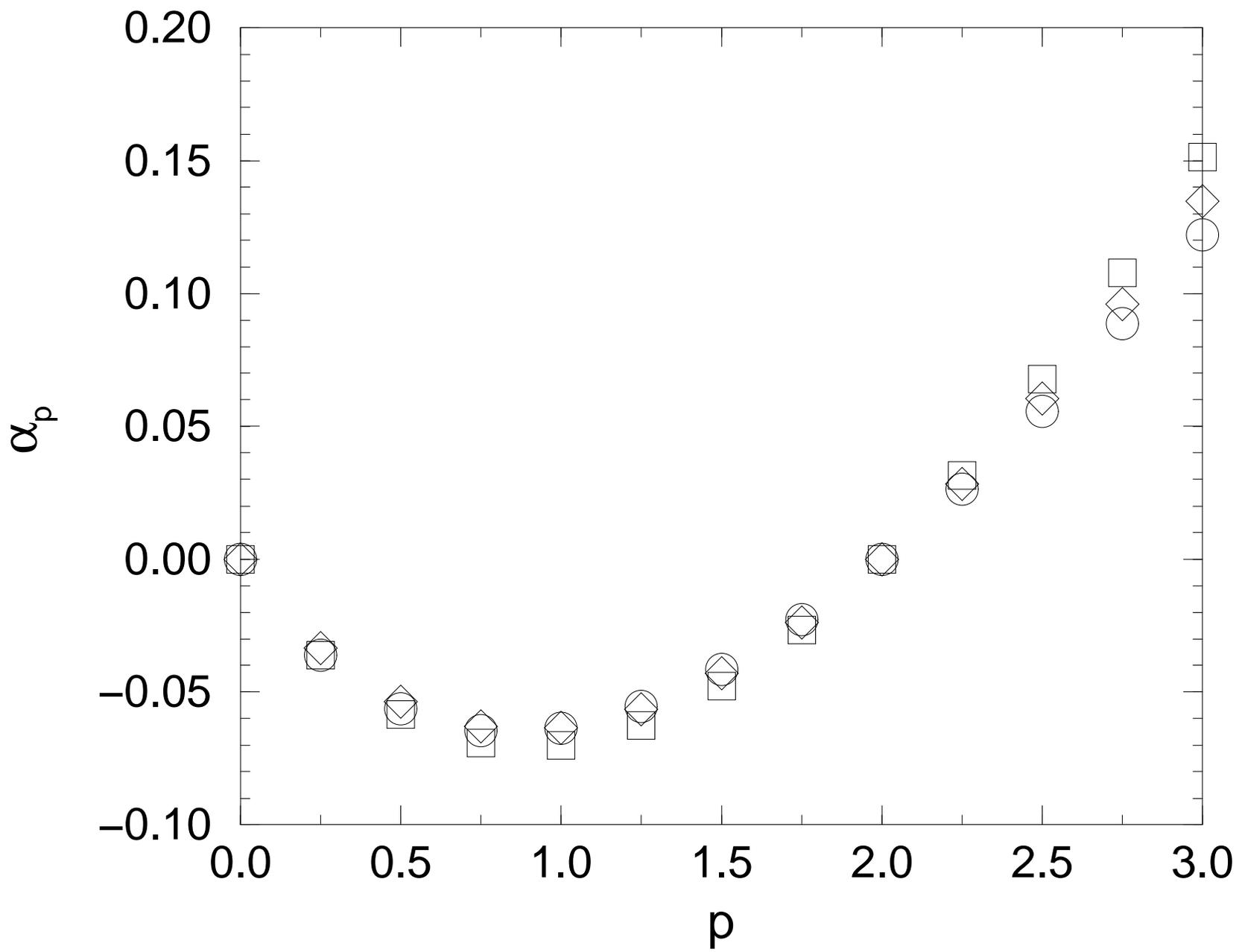
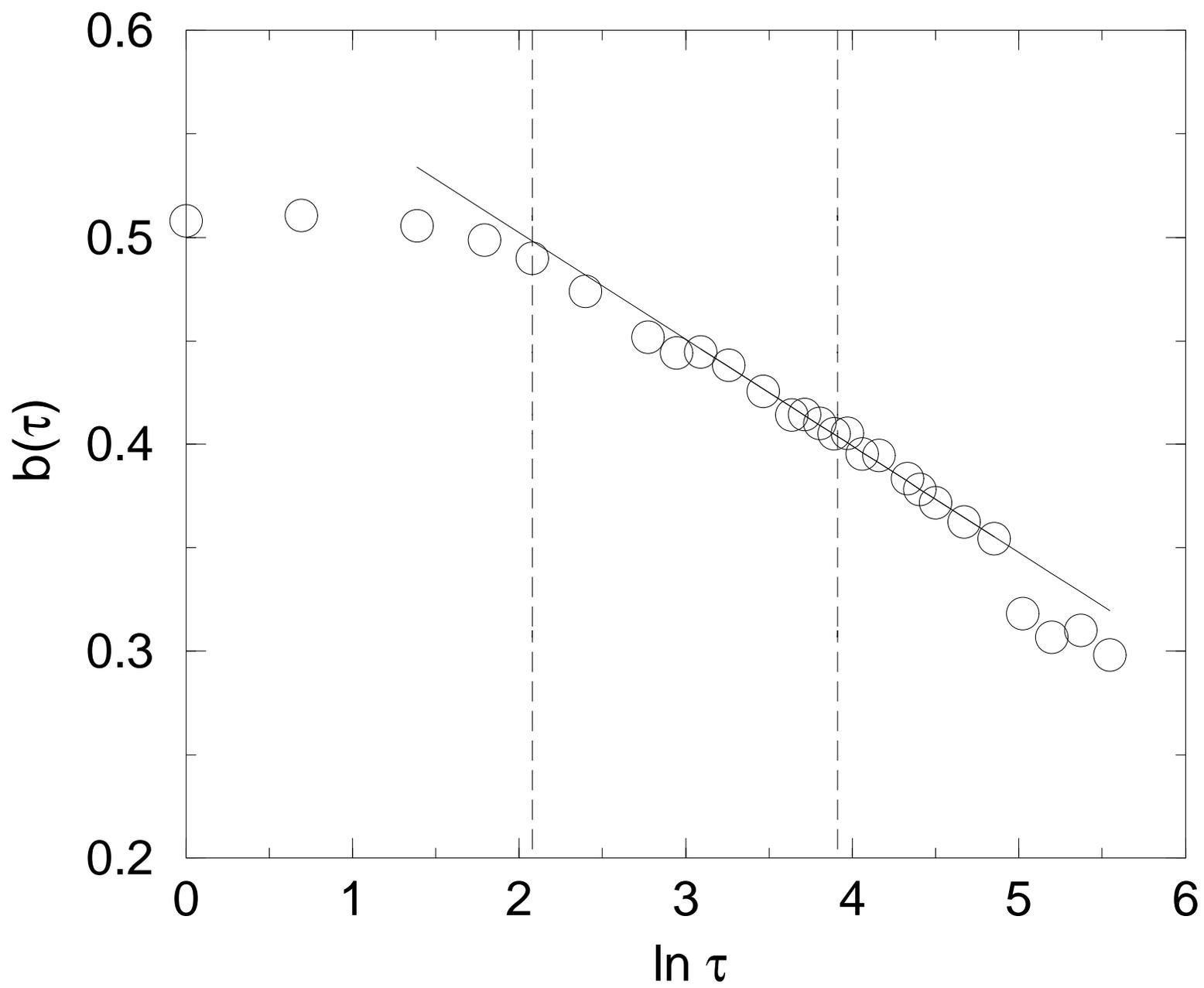
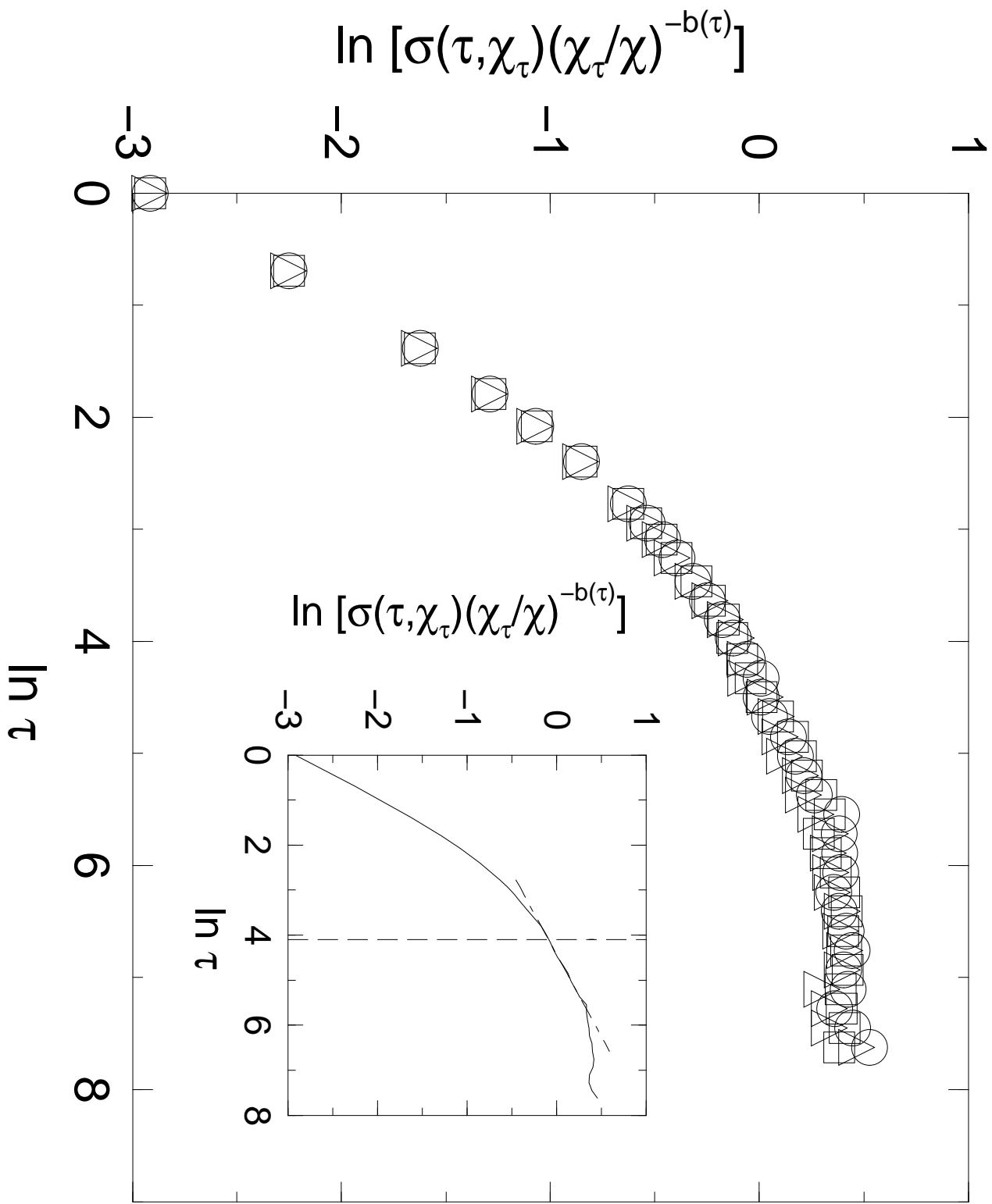
\end{document}